\documentclass[%
 reprint,
 amsmath,amssymb,
 aps,nofootinbib,
]{revtex4-2}

\usepackage{amsmath}
\usepackage{amssymb}
\usepackage{physics}
\usepackage{hyperref}
\usepackage{tikz-feynman}

\usepackage{graphicx}
\usepackage{dcolumn}
\usepackage{bm}

\newcommand{\hc}{\mathrm{h.c.}}
\newcommand{\lagr}{\mathcal{L}}

\begin{document}

\preprint{APS/123-QED}

\title{Higgs Yukawa coupling constraints on a benchmark one-loop\\ radiative mass model for the bottom, charm and tau}

\author{Lucia Stockdale}
    \email{lucia.stockdale@student.unimelb.edu.au}
\author{Raymond R. Volkas}%
 \email{raymondv@unimelb.edu.au}
\affiliation{%
 ARC Centre of Excellence for Dark Matter Particle Physics, School of Physics, \\
The University of Melbourne, Victoria 3010, Australia
}%

\begin{abstract}
Measurements of Higgs boson Yukawa couplings to Standard Model (SM) fermions constrain radiative mass models for those species. Such models, motivated by the observed fermion mass hierarchy, act as foils for the SM tree-level mechanism: we cannot claim to have verified the standard mechanism if other possibilities also fit the data well. We construct a benchmark model which generates the top mass at tree level, and the bottom, charm, tau, and tau Dirac neutrino masses at one-loop level. Current theoretical and experimental constraints on the model, including from Higgs decays, demonstrate it possesses viable regions of parameter space. We show that future improvements to measurements will not be able to rule out the model, only increase the scale of new physics required, illustrating how difficult it will be to verify the SM fermion mass generation mechanism with great precision. As a bonus, a dark matter candidate is shown to be capable of reproducing the correct relic density within the permitted parameter space. 
\end{abstract}

\maketitle

\section{Introduction}

The observed pattern of the Standard Model (SM) fermion masses, and their apparent hierarchical distribution, is a prominent mystery in modern particle physics. While the SM is compatible with the observed masses, and the associated small Yukawa couplings are technically natural, their hierarchical distribution has no apparent dynamical origin. Radiative mass models are often motivated by this issue, and attempt to solve it by generating lighter fermion masses at successively higher loop levels. 

These models have been studied extensively (see~\cite{radiative1,radiative2,radiative3,radiative4,radiative5,radiative6, radiative15,radiative7,radiative8,radiative9,radiative10,radiative11,radiative12,radiative13, Fraser_2014,Fraser:2016pmo,Baker_2021,radiative14} for example). However, few have considered how radiative mass generation modifies the associated Yukawa couplings, an important phenomenological consideration now that Higgs decays to fermions are being measured. In the SM the predicted relationship between a fermion's mass and its Yukawa coupling is
\begin{equation}
    y_f = \frac{\sqrt{2}m_f}{v}
\end{equation}
where $v$ is the Higgs vacuum expectation value (VEV). This relationship is yet to be established at high precision for any of the SM fermions~\cite{ParticleDataGroup:2024cfk}. Existing data constrain the Yukawa couplings for the top, bottom, and charm quarks, the $\tau$, and the $\mu$. Precision on such measurements is expected to increase in the coming decades for the heavier fermions~\cite{collider_future}, with upgrades to the LHC, and prospects for experiments such as the FCC, the ILC, and the CEPC. As this occurs, it becomes important to examine how alternative models of mass generation modify (or replicate) this relationship. They can be constrained by such measurements if no disagreement with the SM is found, or used to explain such a disagreement if it is measured.

\vspace{-0.01cm}
Another aspect that has not been extensively studied is the generation of \textit{all} fermion masses, except for that of the top quark, at loop level; see Refs.~\cite{radiative4,radiative5,radiative6, radiative9, radiative10,Baker_2021} for existing analyses of such scenarios. The motivation is simply that the top quark mass is the only fermion mass close to the electroweak (EW) scale, so is naturally expected to arise at tree-level. All of the others are suppressed, plausibly because they arise at loop level. However, many radiative models take the bottom quark and tau lepton masses to be generated at tree-level along with the top mass. 

\begin{table*}[t]
\caption{\label{tab:c1aqn}Quantum numbers of relevant fields in our model, including extended lepton number $L$ and extended baryon number $B$. The $SU(3)_C$ labels are the only allowable ones using fundamental or singlet representations. The $SU(2)_L$ labels are chosen to give the colour singlet exotics integer electric charge. See the main text for an explanation of the role played by each of the non-SM symmetries.}
\begin{ruledtabular}
\begin{tabular}{c|ccccccccccccc}
\multicolumn{1}{l|}{} & $\tau_R$ & $\nu_R$ & $L_L$          & $b_R$           &$c_R$& $Q_L$         & $H$           & $\psi_L$       & $\psi_R$       & $\chi_L$       & $\chi_R$       & $\phi$         & $\eta$         \\ \hline
$SU(3)_C$             & $\mathbf{1}$& $\mathbf{1}$& $\mathbf{1}$& $\mathbf{3}$ &$\mathbf{3}$ & $\mathbf{3}$& $\mathbf{1}$& $\mathbf{1}$& $\mathbf{1}$& $\mathbf{3}$& $\mathbf{3}$& $\mathbf{1}$& $\mathbf{1}$\\
$SU(2)_L$             & $\mathbf{1}$& $\mathbf{1}$& $\mathbf{2}$& $\mathbf{1}$ &$\mathbf{1}$ & $\mathbf{2}$& $\mathbf{2}$& $\mathbf{2}$& $\mathbf{2}$& $\mathbf{2}$& $\mathbf{2}$& $\mathbf{1}$& $\mathbf{2}$\\
$U(1)_Y$              & $-1$     & $0$     & $-\frac{1}{2}$ & $-\frac{1}{3}$  &$\frac{2}{3}$& $\frac{1}{6}$ & $\frac{1}{2}$ & $-\frac{1}{2}$ & $-\frac{1}{2}$ & $\frac{1}{6}$ & $\frac{1}{6}$ & $0$& $\frac{1}{2}$ \\ \hline
$L$                   & $1$      & $1$     & $1$            & $0$             &$0$             & $0$           & $0$           & $1$           & $1$           & $0$           & $0$           & $0$           & $0$           \\
$3B$                  & $0$      & $0$     & $0$            & $1$             &$1$             & $1$           & $0$           & $0$           & $0$           & $1$           & $1$           & $0$           & $0$           \\
$\mathcal{Z}_2$       & $1$      & $1$     & $1$            & $1$             &$1$             & $1$           & $1$           & $-1$           & $-1$           & $-1$           & $-1$           & $-1$           & $-1$           \\ \hline
$U(1)_a$& $1$& $-1$& $0$            & $1$ &$-1$& $0$           & $0$           & $0$            & $0$            & $0$            & $0$            & $0$            & $-1$           \\
$U(1)_\psi$& $1$      & $1$            & $0$            & $0$             &$0$             & $0$           & $0$           & $1$            & $0$            & $0$            & $0$            & $0$            & $0$            \\
$U(1)_\chi$& $0$            & $0$            & $0$            & $1$             &$1$             & $0$           & $0$           & $0$            & $0$            & $1$            & $0$            & $0$            & $0$           
\end{tabular}
\end{ruledtabular}
\end{table*}

\vspace{-0.01cm}
The purpose of this paper is to progress the study of one-loop radiative models for the bottom, tau and charm masses in light of Higgs Yukawa coupling constant and other constraints. These species are selected because they are all at the few-GeV level as would be expected from a $1/16\pi^2$ one-loop suppression factor. The benchmark model we present also produces a neutrino Dirac mass at the same loop level. The physical neutrino mass then must arise from further suppression, for example through a seesaw effect, though we do not consider any specific implementation. While we discuss a specific model in this paper, we also intend to present a systematic catalogue of related models in a companion paper. We anticipate that our work will form the basis of extensions to higher loop radiative models for the generation of the smaller fermion masses. 

\vspace{-0.01cm}
We build on the work of Refs.~\cite{Fraser_2014,Fraser:2016pmo,Baker_2021} which explored the modified Yukawa couplings when the $b$ quark and $\tau$ masses are generated at one-loop, seeded by soft-breaking parameters for symmetries that forbid their tree-level masses. As stated above, we extend these frameworks to also produce one-loop charm and neutrino Dirac masses. We do so in light of recent updates to measurements, and from the perspective of signals at colliders. We examine the current constraints and future sensitivities on this model for selected slices of parameter space, and briefly outline opportunities for future research. We also extend Refs.~\cite{Fraser_2014,Fraser:2016pmo,Baker_2021} by demonstrating that our benchmark model contains a viable dark matter candidate, a fairly common feature in this class of theories, which adds to their importance.

In Sec.~\ref{sec:model} we define the model, we compute current phenomenological constraints and future sensitivities in Sec.~\ref{sec:pheno}, with the viability of the dark matter candidate established in Sec.~\ref{sec:dm}. We end with discussions of future work and concluding remarks in Secs.~\ref{sec:futurework} and~\ref{sec:conclusion}.

\section{Model}
\label{sec:model}

Our model is built using the framework presented in~\cite{Baker_2021}, but extended to treat upper components of electroweak doublets. We introduce four exotic fields to the theory: two vectorlike fermions $\psi$ and $\chi$, a complex scalar $\eta$, and a real scalar $\phi$. The relevant Lagrangian is
\begin{eqnarray}
        -\lagr \supset &&y_L^L \overline{L_L} \phi \psi_R + y_L^{Q_i} \overline{Q_L^{i}} \phi \chi_R  + y_R^\tau \overline{\psi_L} \eta \tau_R \nonumber \\
        &&+ y_R^\nu \overline{\psi_L} \Tilde{\eta} \nu_R + y_R^b \overline{\chi_L} \eta b_R  + y_R^c \overline{\chi_L} \Tilde{\eta} c_R \nonumber \\
        &&+ aH \eta^\dagger \phi + m_\psi \overline{\psi_L} \psi_R + m_\chi \overline{\chi_L} \chi_R + \hc \, 
\end{eqnarray}
where $\Tilde{\eta} = i \sigma_2 \eta^*$, $L_L$ is the third generation lepton doublet, and $Q_L^{2(3)}$ is the second (third) generation quark doublet. We assume $\eta$ does not acquire a VEV.

The simplest quantum number assignments for the relevant fields given this Lagrangian are shown in Table~\ref{tab:c1aqn}. We remain agnostic as to the structure of a full theory treating all three generations of quarks and leptons, and thus do not consider any terms coupling exotics to the remaining SM fermions in our analysis.

This Lagrangian admits three softly broken symmetries, denoted by $U(1)_a$, $U(1)_\psi$, and $U(1)_\chi$. The first of these is softly broken by the trilinear term, and $U(1)_\psi$ and $U(1)_\chi$ are softly broken by the $\psi$ and $\chi$ mass terms respectively. The tree level Yukawa terms for $b$ and $c$ will be hard-breaking terms for $U(1)_{a,\chi}$, while the $\tau$ and $\nu_\tau$ tree level Yukawa terms will be hard-breaking terms for $U(1)_{a, \psi}$, and thus all four terms are forbidden. None of these three $U(1)$s can be imposed on the theory as exact symmetries, since this would prevent at least one SM fermion from gaining mass at any order. We do, however, impose exotic particle parity $\mathcal{Z}_2$ (defined in Table~\ref{tab:c1aqn}) as an exact symmetry of the theory, to prevent mixing between exotics and SM fields. As a result, the lightest exotic will be stable, and thus a dark matter candidate if it is colourless and electrically neutral. We restrict our analysis to regions of parameter space where the lightest exotic satisfies these conditions, and all other exotics can decay into it along with some combination of SM fields.

After electroweak symmetry breaking (EWSB), $\phi$ will mix with the real part of the lower component $\eta^{low}$ of $\eta$, with mixing angle $\theta_s$, working in the basis in which $\Im{\eta^{low}}$ does not mix with $\phi$ and $\Re{\eta^{low}}$ at tree level. We denote the mass eigenstate superpositions of $\phi$ and $\Re{\eta^{low}}$ as $\varphi_{1,2}$, with $\varphi_1$ being lighter. The resulting scalar mass eigenvalues are
\begin{equation}
    m_{1,2}^2 = \frac{1}{2} \left( m_\phi^2 + m_\eta^2 \pm \sqrt{\left(m_\eta^2 - m_\phi^2  \right)^2 + 4 a^2v^2} \right)~.
\end{equation}
We also obtain the relationship
\begin{equation}
    \sin 2\theta_s = \frac{2 av}{m_{2}^2 - m_{1}^2} \, .
\end{equation}
These results differ from those in~\cite{Baker_2021} only by certain factors of two, which arise from real rather than complex scalars mixing. 

For ease of notation, we also define
\begin{equation}
    \alpha = \left\{
      \begin{array}{ll}
        \psi, & \text{for } f=\tau,\nu_\tau \\
        \chi, & \text{for } f=b,c.
      \end{array}
    \right.
\end{equation}

\begin{figure*}
    \begin{minipage}[b]{0.49\textwidth}
        \centering
        \begin{tikzpicture}
        \begin{feynman}
\vertex (a2);
\vertex[left=1.5cm of a2] (a1) {$f_R$};
\vertex[right=2.4cm of a2] (a3);
\vertex[right=1.5cm of a3] (a4) {$f_L$};
\vertex[right=1.2 cm of a2] (fill);
\vertex[above=1.2 cm of fill] (d);
\vertex[above=0.01 cm of d] (lab) {$\varphi_{1,2}$};

\diagram* {
(a1) -- [fermion] (a2) -- [scalar, quarter left] (d) -- [scalar, quarter left] (a3) -- [fermion] (a4),
(a2) -- [with arrow = 0.25, insertion = 0.5, with arrow = 0.75, edge label'={\vspace{0.1cm}$\alpha$}] (a3) ,
};

\end{feynman}
        \end{tikzpicture}
    \end{minipage}
    ~ 
    \begin{minipage}[b]{0.49\textwidth}
        \centering
        \begin{tikzpicture}
        \begin{feynman}
\vertex (a2);
\vertex[left=1.5cm of a2] (a1) {$f_R$};
\vertex[right=2.4cm of a2] (a3);
\vertex[right=1.5cm of a3] (a4) {$f_L$};
\vertex[right=1.2 cm of a2] (fill);
\vertex[above=1.2 cm of fill] (d);
\vertex[above=1 cm of d] (lab) {$h$};

\diagram* {
(a1) -- [fermion] (a2) -- [scalar, quarter left, edge label=$\varphi_{1,2}$] (d) -- [scalar, quarter left, edge label = $\varphi_{1,2}$] (a3) -- [fermion] (a4),
(d) -- [scalar] (lab),
(a2) -- [with arrow = 0.25, insertion = 0.5, with arrow = 0.75, edge label'={$\alpha$}] (a3) ,
};

\end{feynman}
        \end{tikzpicture}
        \vspace{0.15cm}
    \end{minipage}
    \caption{\label{fig:feynmandiagrams}Mass generation (left) and effective Yukawa coupling (right) diagrams for $f=\tau,\nu_\tau,b,c$ at one loop after EWSB. For $\tau$ and $\nu_\tau$ the virtual fermion $\alpha$ is $\psi$, while $\alpha = \chi$ for $b$ and $c$.}
\end{figure*}
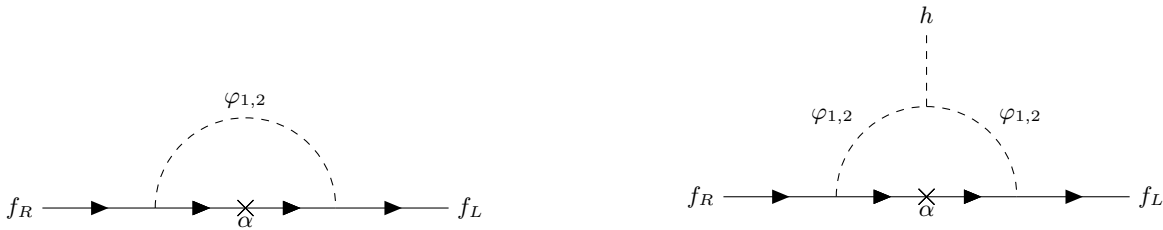

The Feynman diagrams for SM fermion masses and Yukawa couplings after EWSB are given in Figure~\ref{fig:feynmandiagrams}. Our expressions for the masses and Yukawa couplings are identical to those in~\cite{Baker_2021}, 
\begin{eqnarray}\label{eq:mandy}
        y_{f}^\mathrm{eff}|_{p_h^2 = 0} &=& \frac{y_L^f y_R^{f}}{16\pi^2}  \frac{a m_\alpha}{m_{1} m_{2}} \left[\cos^2 2\theta_s\  F \left( \frac{m_{1}^2}{m_\alpha^2}, \frac{m_{2}^2}{m_\alpha^2} \right) \right. \nonumber \\
        &&\left.+ \frac{\sin^2 2\theta_s}{2} \left( \frac{m_{2}}{m_{1}} F \left( \frac{m_{1}^2}{m_\alpha^2} \right) + \frac{m_{1}}{m_{2}} F \left( \frac{m_{2}^2}{m_\alpha^2} \right)  \right) \right]~, \nonumber \\
        m_{f} &=& \frac{v}{\sqrt{2}}\frac{y_L^f y_R^f}{16\pi^2} \frac{a m_\alpha}{m_{2} m_{1}} F\left( \frac{m_{_1}^2}{m_\alpha^2}, \frac{m_{_2}^2}{m_\alpha^2} \right)~,
\end{eqnarray}
where we have defined
\begin{eqnarray}
        &&F(x,y) = \frac{\sqrt{xy}}{x-y} \left(\frac{x}{x-1} \ln \left(x\right) - \frac{y}{y-1} \ln \left(y\right) \right)~, \nonumber \\
        &&F(x) = \lim_{y\rightarrow x} F(x,y) =\frac{x}{x-1} - \frac{x}{(x-1)^2} \ln \left(x \right)~.
\end{eqnarray}
The full expression for $y_f^\mathrm{eff}$ is unwieldy, so we have provided it here in the $m_h \rightarrow 0$ limit, but use the full version for our plots. Since $F(x,y) > 0 \ \forall x,y>0$, we have $y_f^\mathrm{eff} > y_f^\mathrm{SM}$ in this model.

The $t$ and $c$ quarks will mix at one-loop, since $t_L$ couples to the exotics via the $y_L^{Q_3}$ term. The effective Yukawa and mass terms at one loop are
\begin{eqnarray}
    -\lagr &\supset& \mqty(\overline{t_L} & \overline{c_L}) \mqty(\frac{y_t v}{\sqrt{2}} & \frac{y_L^{Q_3}}{y_L^{Q_2}}m_c \\ \frac{(y_L^{Q_3} y_R^c)^*}{y_L^{Q_2} y_R^c} m_c & m_c) \mqty(t_R \\ c_R) +\hc \\
    &&+ \mqty(\overline{t_L} & \overline{c_L}) \mqty(y_t& \frac{y_L^{Q_3}}{y_L^{Q_2}} y_c^\mathrm{eff} \\ \frac{(y_L^{Q_3} y_R^c)^*}{y_L^{Q_2} y_R^c} y_c^\mathrm{eff} & y_c^\mathrm{eff}) \mqty(t_R \\ c_R)\frac{h}{\sqrt{2}} + \hc \nonumber
\end{eqnarray}
where $m_c, y_c^\mathrm{eff}$ are as defined in eq.~\ref{eq:mandy}. In general, these will not be simultaneously diagonalisable, and thus there will exist an interaction between $t$, $c$, and $h$, in the mass basis. Diagonalising the mass matrix is trivial, so we refrain from displaying the unwieldy formulas for the eigenvalues and mixing angles. 
In the mass basis, we rewrite the effective Yukawa coupling terms for $t$ and $c$ as
\begin{equation}
    - \lagr \supset \mqty(\overline{t_L'} & \overline{c_L'}) Y_\mathrm{eff} \mqty(t_R' \\ c_R') \frac{h}{\sqrt{2}} + \hc~.
\end{equation}
Taking $y_L^3 = y_R^b$ and $y_L^2 = y_R^c$ to obtain indicative values, we find
\begin{equation}\label{eq:offdiag}
    Y_\mathrm{eff} = \mqty(0.991 & 0.014 \\ 0.014 & 0.00018) +  \mqty(0.015 & -0.563 \\ -0.563 & 0.297 ) y_b^\mathrm{eff}
\end{equation}
where $y_b^\mathrm{eff}$ is defined in eq.~\ref{eq:mandy}. The partial decay width and branching ratio for the novel $t\rightarrow ch$ decay depends on the off-diagonal elements of $Y_\mathrm{eff}$.

\section{Phenomenology}
\label{sec:pheno}

\begin{figure*}
\centering
    \begin{minipage}{0.49\linewidth}
        \centering
        \includegraphics[width=6.5cm]{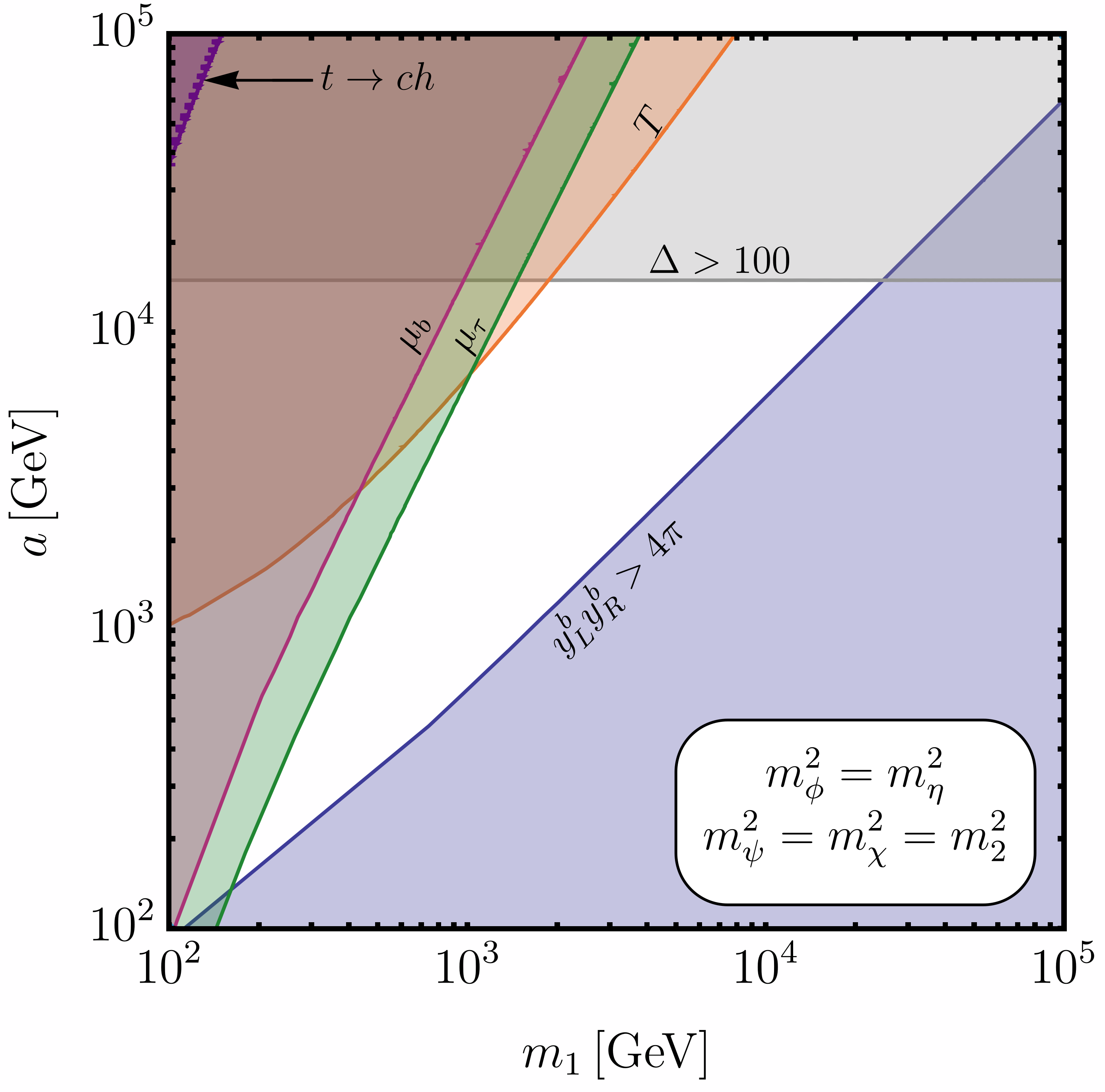}
    \end{minipage}
    \hfill
    \begin{minipage}{0.49\linewidth}
        \centering
        \includegraphics[width=6.5cm]{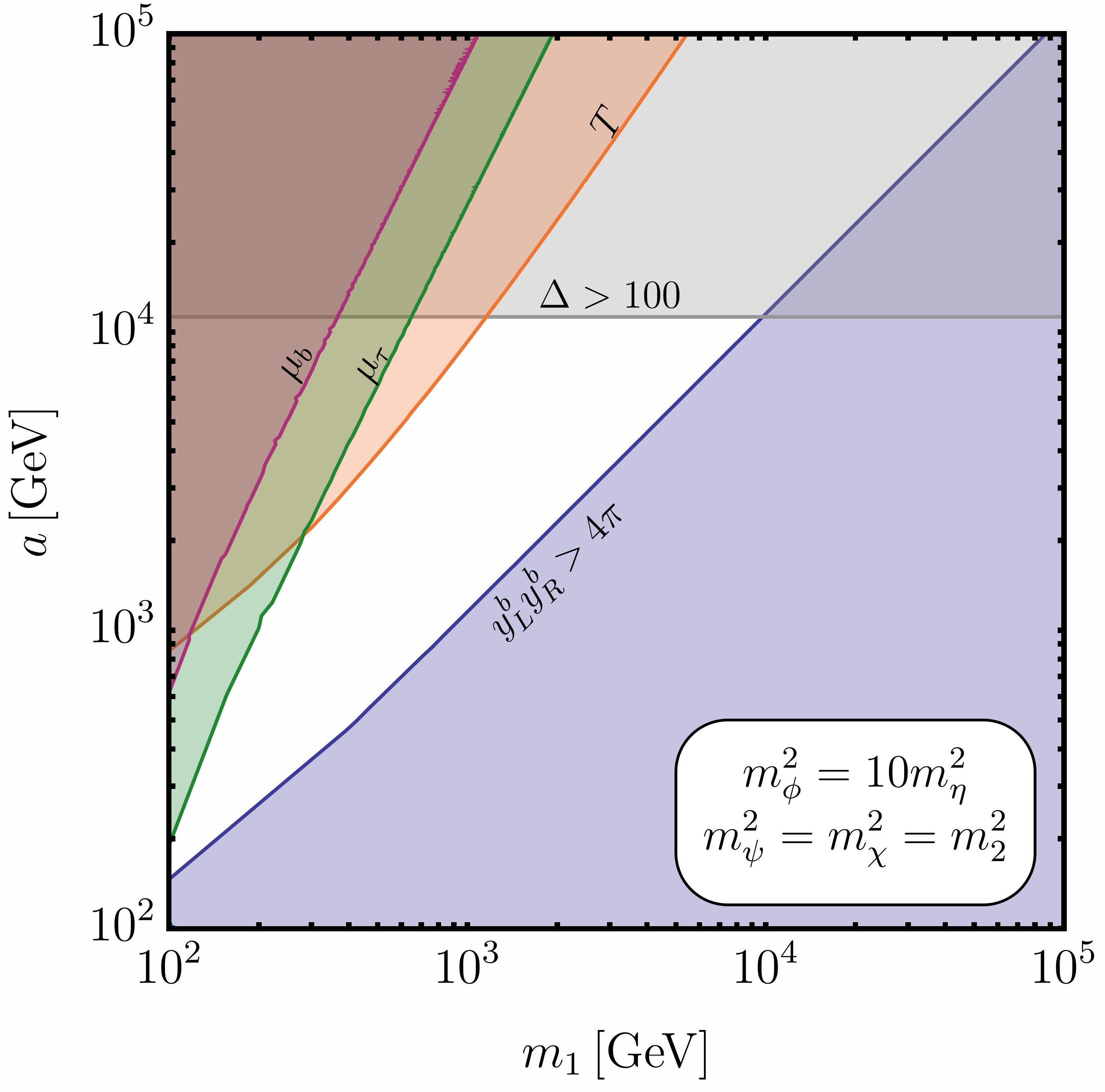}
    \end{minipage}
\caption{\label{fig:constraints}Theoretical and current $2\sigma$ experimental constraints on the model for two benchmark slices of parameter space. Shaded regions are excluded, except for the case of $\Delta > 100$ where they are disfavoured.}
\end{figure*}

We considered a variety of constraints to determine whether this model is phenomenologically viable. The theoretical constraints we examine are the perturbative unitarity of the various $y_{L,R}$, and fine-tuning for the Higgs mass parameter $\mu_H^2$. Specifically, for the former, we can calculate the product $y_L^f y_R^f$ by requiring the observed fermion mass for $f$ be reproduced, and then exclude $y_L^f y_R^f > 4\pi$ on our plots. The constraint on the $b$ couplings is always strongest for the slices of parameter space shown here, so we do not explicitly show constraints on the other couplings. We adopt the fine-tuning measure $\Delta$ defined in~\cite{finetune}, and take $\Delta > 100$ to be our benchmark for a high degree of fine-tuning. For consistency with~\cite{Baker_2021}, we take the matching scale to be $\sqrt{m_\phi m_\eta}$. While regions with $\Delta > 100$ may be technically permitted, introducing an unnatural hierarchy between the electroweak and new physics scales undermines the motivation of explaining the hierarchy of fermion masses. We do not show the full perturbative unitarity constraint on the trilinear coupling constant $a$, instead noting that it results in an upper bound on $a$ roughly of the form $\mathcal{O}(1) \cdot m_1$. Indicative perturbativity bounds on $a$ can be found in~\cite{Baker_2021}.

\begin{table}[b]
\caption{\label{tab:signals}Signal strengths for Higgs boson decays into relevant SM fermions~\cite{ParticleDataGroup:2024cfk}, normalised to unity at the SM prediction. The dominant contributions to these values are from ATLAS and CMS data~\cite{ATLAS_tau,ATLAS_bb,CMS_sum,CMS_cc}.}
\begin{ruledtabular}
\begin{tabular}{c|ccc}
    Process&   $h\rightarrow \tau^+ \tau^-$& $h \rightarrow \overline{b} b$&$h \rightarrow \overline{c} c$\\
     \hline
     $\mu_f$ & $0.91\pm0.09$& $0.94\pm0.11$&$-0.5 \pm 3.4$
\end{tabular}
\end{ruledtabular}
\end{table}

The experimental constraints we consider are the most recent bounds on the signal strengths for Higgs decays to the relevant fermions, modifications to EW physics, and the novel $t\rightarrow ch$ decay. The signal strength for a process $ii~\rightarrow~h~\rightarrow~ff$ is defined in the $\kappa$ formalism~\cite{kappaframe,kappaframelong} as
\begin{equation}
    \mu_f^i = \frac{\kappa_i^2 \kappa_f^2}{\kappa_h^2}
\end{equation}
All signal strengths are normalised to unity at the SM prediction. In this model there is no modification (at tree or one-loop level) to the Higgs boson coupling to gluons or photons, and a negligible modification to its coupling to the $W$ and $Z$ bosons, so we consider only the modified Yukawa couplings and Higgs width. The current constraints on the relevant signal strengths are shown in Table~\ref{tab:signals}. Note that for $m_\psi = m_\chi$, our model predicts $\kappa_\tau = \kappa_b = \kappa_c$ and thus $\mu_\tau = \mu_b = \mu_c$, so the strongest constraint will always be provided by the signal strength with the most precise measurement.

\begin{figure*}
    \centering
    \begin{minipage}{0.49\linewidth}
        \centering
        \includegraphics[width=6.5cm]{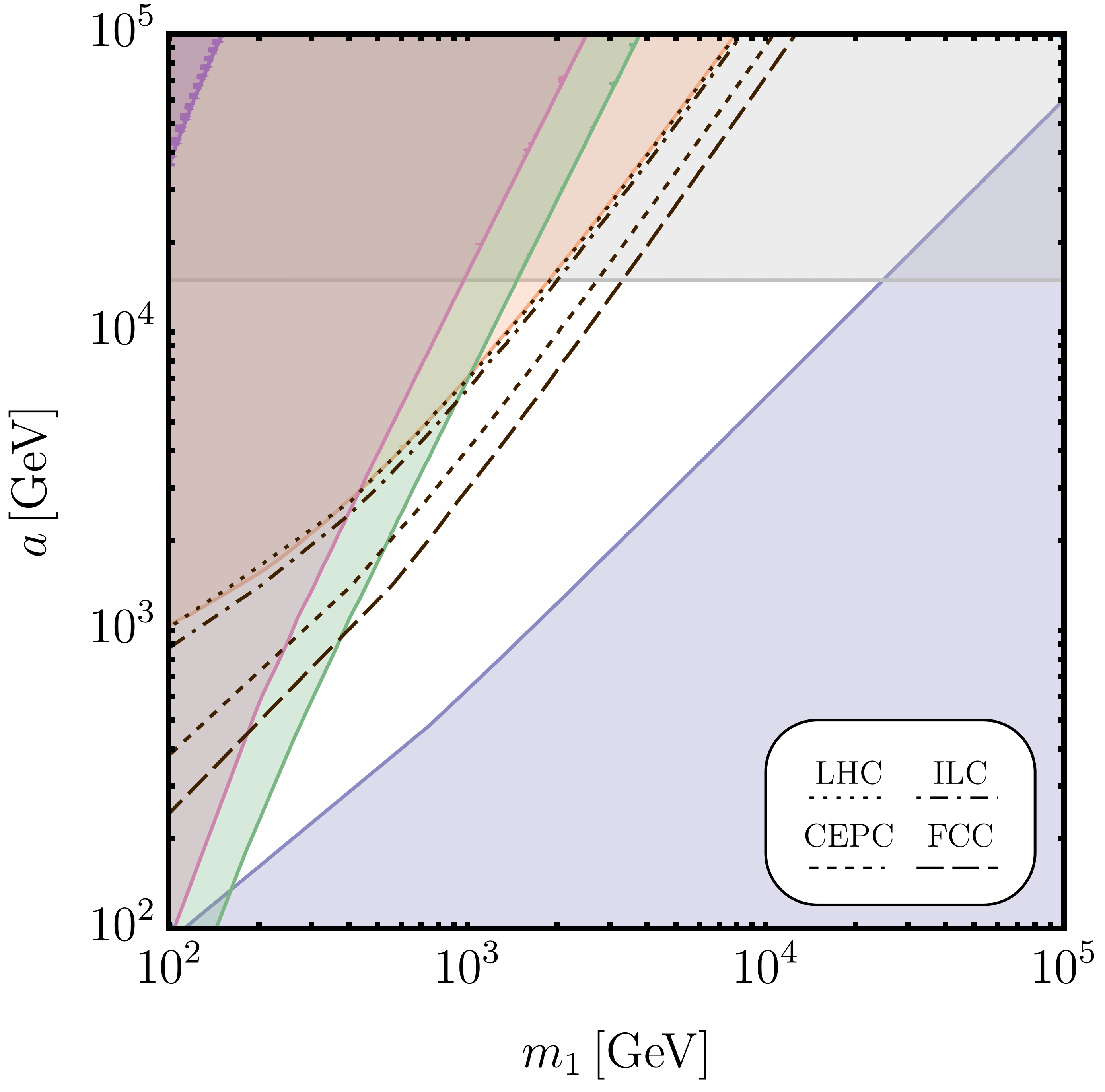}
    \end{minipage}
    ~
    \begin{minipage}{0.49\linewidth}
        \centering
        \includegraphics[width=6.5cm]{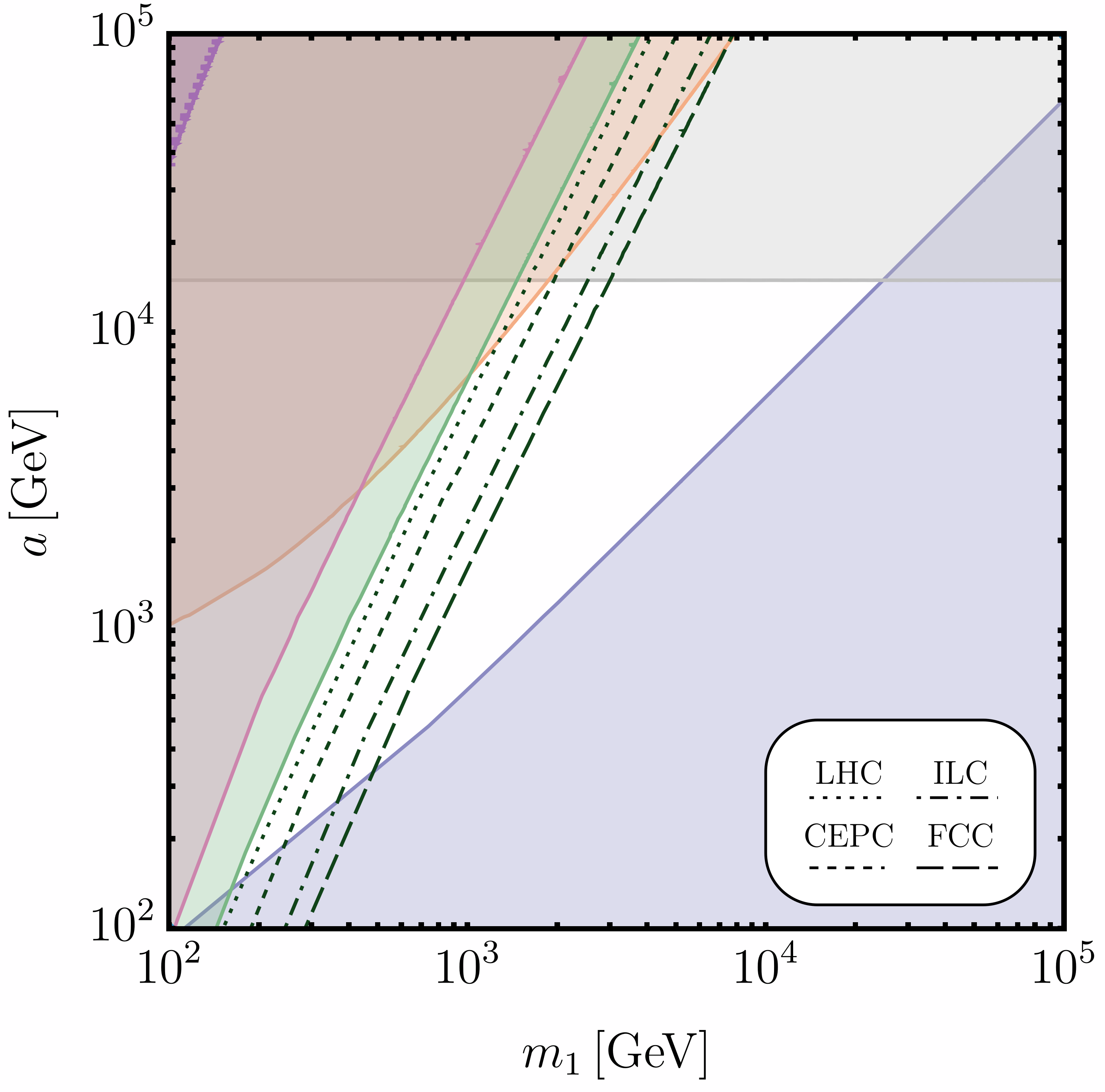}
    \end{minipage}
    \vspace{0.4cm}
    \begin{minipage}{0.49\linewidth}
        \centering
        \includegraphics[width=6.5cm]{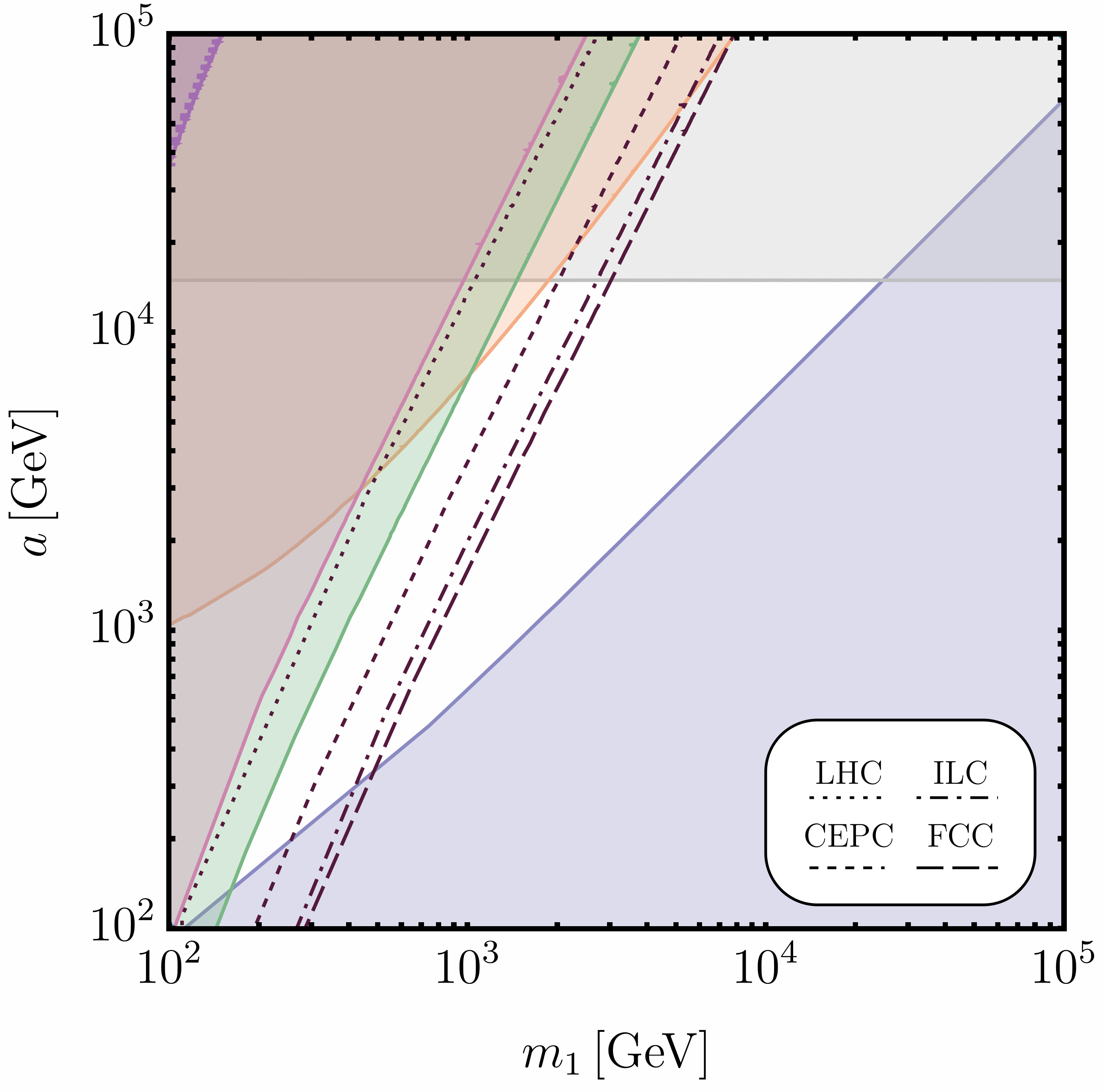}
    \end{minipage}
    ~
    \begin{minipage}{0.49\linewidth}
        \centering
        \includegraphics[width=6.5cm]{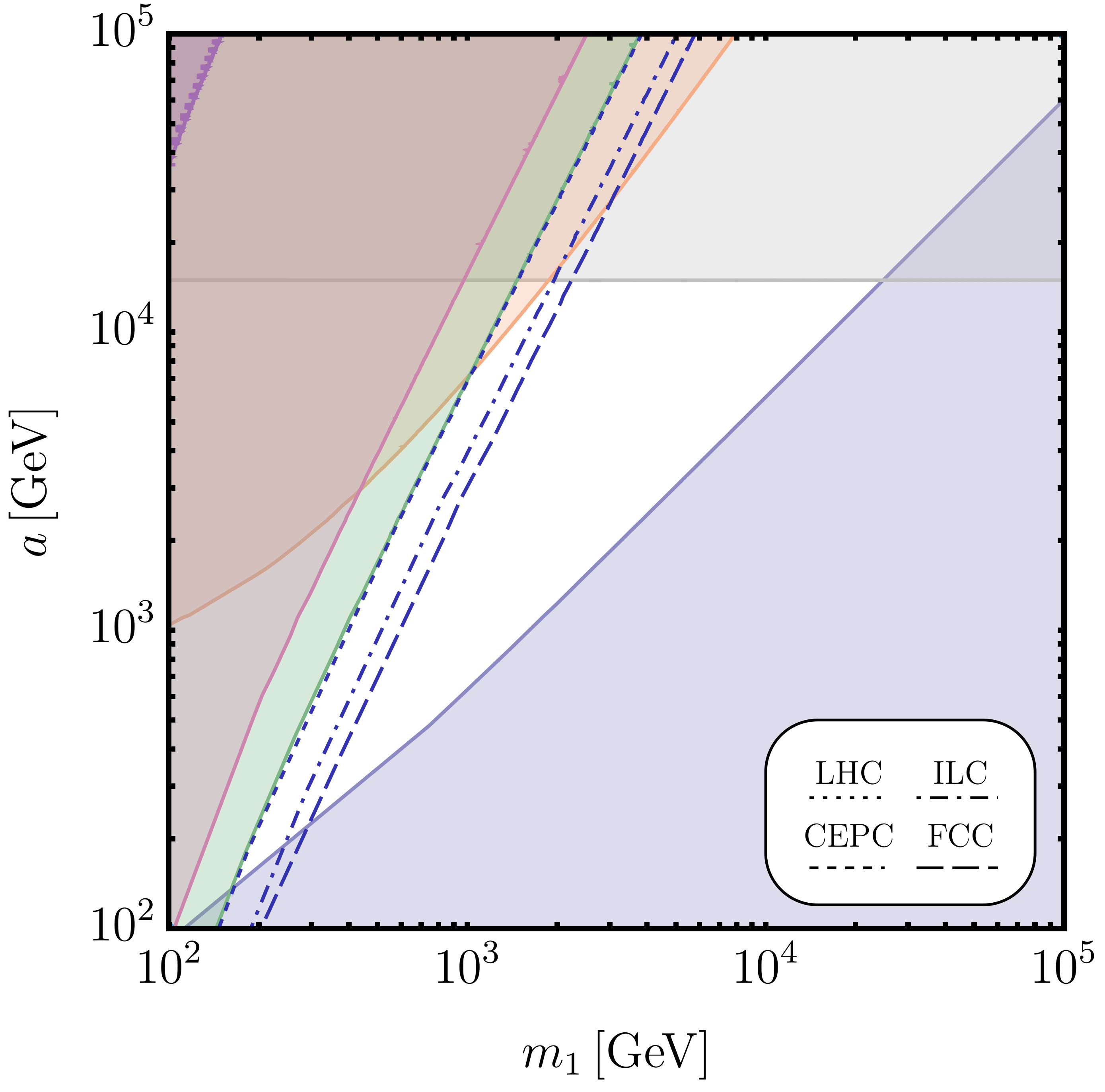}
    \end{minipage}
    \caption{\label{fig:future}Projected $2\sigma$ experimental sensitivities when $m_\phi^2 = m_\eta^2$ and $m_\psi^2 = m_\chi^2 = m_2^2$. Plots shown are the sensitivities for $T$ (top left), $\mu_\tau$ (top right), $\mu_b$ (bottom left), and $\mu_c$ (bottom right). Values for $\Delta T$ taken from~\cite{STfuture} and for $\Delta\kappa_\tau, \Delta\kappa_b, \Delta\kappa_c$ from~\cite{collider_future}. Shaded regions correspond to the same constraints as in Figure~\ref{fig:constraints}.}
\end{figure*}

Corrections to EW physics will primarily occur through contributions to the vacuum polarisations of EW gauge bosons, rather than box or vertex corrections, justifying use of the Peskin-Takeuchi parameters~\cite{PTparamsearly,PTparams}. We only consider corrections to the $S$ and $T$ parameters, ignoring any modifications to $U$ since these are typically smaller. Only new physics contributions are considered, so $S=T=U=0$ is the SM prediction. Under the assumption that $U=0$, the current $1\sigma$ constraints are~\cite{ParticleDataGroup:2024cfk}
\begin{eqnarray}
        S &=& 0.008 \pm 0.071 \nonumber \\
        T &=& 0.021 \pm 0.055~.
\end{eqnarray}
Neither $\psi$ nor $\chi$ will provide any net contribution to these. We find that $T$ provides a stronger bound than $S$ in all regions of parameter space we explore, so we explicitly display the formula and bounds for $T$ only. The one-loop scalar contributions to $T$ are
\begin{eqnarray}
    T = \left[\frac{2 s_s^2 c_s^2 m_1^2 m_2^2}{m_1^2 - m_2^2} \ln \left(\frac{m_1^2}{m_2^2} \right) + m_1^2 c_s^4 + m_2^2 s_s^4 +m_\eta^2 \right.&& \nonumber \\
    + \left. \frac{2 c_s^2 m_1^2 m_\eta^2}{m_1^2 - m_\eta^2} \ln \left(\frac{m_\eta^2}{m_1^2} \right) + \frac{2 s_s^2 m_2^2 m_\eta^2}{m_2^2 - m_\eta^2} \ln \left(\frac{m_\eta^2}{m_2^2} \right) \right]&&\\
    \times \frac{1}{16 \pi c_W^2 s_W^2 m_Z^2}&& \nonumber
\end{eqnarray}
where we have denoted $\sin \theta_i = s_i$ and $\cos \theta_i = \theta_i$.

The rare $t\rightarrow ch$ decay will be induced by mixed interactions of $t$ and $c$. The partial decay width for $t\rightarrow ch$ is
\begin{eqnarray}
    \Gamma_{t\rightarrow ch} =  \frac{m_t}{32\pi} \left(1- \frac{ m_h^2}{m_t^2} \right)^2 (0.014 - 0.563 y^\mathrm{eff}_b)^2
\end{eqnarray}
using the values from eq.~\ref{eq:offdiag}. We can compare this to the partial decay width
\begin{equation}
    \Gamma_{t\rightarrow bW} =\frac{|V_{tb}|^2}{16\pi} \frac{m_t^3}{v^2}\left(1+2 \frac{m_W^2}{m_t^2}  \right) \left(1-\frac{ m_W^2}{m_t^2} \right)^2
\end{equation}
Assuming that $Br(t\rightarrow bW)$ is approximately unity, we write
\begin{equation}
    Br(t\rightarrow ch) \simeq \frac{\Gamma_{t\rightarrow ch}}{\Gamma_{t\rightarrow bW}} = 0.261 \cdot \left(0.014 - 0.563 y^\mathrm{eff}_b \right)^2
\end{equation}
Similar to $\mu_b$, this expression has a quadratic dependence on $y_b^\mathrm{eff}$, so these bounds will have a similar form. The current $2\sigma$ upper bound on $Br(t\rightarrow ch)$ is $3.4 \cdot 10^{-4}$~\cite{tchbound}, though it could be brought as low as $1.6 \cdot 10^{-5}$ with the next generation of experiments~\cite{ESUbriefing}. We find that the projected bounds on the branching fraction are weaker than the current bound on $\mu_\tau$ for the parameter space considered, and thus do not show them here.

Current $2\sigma$ constraints on this model are shown for two benchmark slices of parameter space in Figure~\ref{fig:constraints}. As can be seen from these plots, viable parameter space with $<1\%$ fine-tuning currently exists for this model. This indicates that it remains sensible to use this model as a foundation for constructing a complete model. The most stringent constraints are generally perturbativity, the $\mu_\tau$ signal strength, and corrections to $T$.

Bounds obtained from projected future sensitivities~\cite{collider_future,STfuture} for $\mu_f$ and $T$ are shown in Figure~\ref{fig:future}, assuming agreement with the SM is maintained. Assumptions on the integrated luminosities and centre of mass energies for these projections are given in Table~1 of~\cite{collider_future}, where we have considered the HL-LHC, ILC\textsubscript{500}, the full CEPC program, and the full FCC-ee/eh/hh program in our plots.

Another avenue for constraining such a model is the bounds obtained from direct searches at colliders. Since $\chi$ is the only coloured exotic, we would expect the bound on its mass to be strongest. The relevant searches are for vector-like quarks (VLQs). However none of the existing VLQ searches can be repurposed, as they assume that the given VLQ primarily decays into SM fields~\cite{VLQs}, something $\chi$ cannot do by the imposition of exotic $\mathcal{Z}_2$. Future searches for VLQs that relax this assumption, or even consider VLQs that cannot decay into exclusively SM fields, would thus be helpful for constraining models that contain fields similar to $\chi$. Existing bounds on squarks and VLQs suggest a lower bound of $m_\chi \gtrsim 1 \ \mathrm{TeV}$ would be typical~\cite{VLQrev,squark1,squark2}. 

To repurpose bounds from direct searches at colliders for other fields, we note that $\phi$ resembles a right-handed anti-sneutrino, $\eta$ resembles a left-handed anti-slepton doublet, and $\psi$ resembles an anti-higgsino. Assuming $m_\psi \ll m_\eta$, the lower bound on the $\eta$ mass is $m_\eta \gtrsim 425$ GeV~\cite{SUSYbound} This bound gets weaker as $m_\psi$ increases, and there is no constraint for $m_\psi \gtrsim 140 \ \mathrm{GeV}$. This bound thus does not exclude any significant region of parameter space.

\section{Dark matter}
\label{sec:dm}

\begin{figure}[t]
    \centering
    \includegraphics[width=6.5cm]{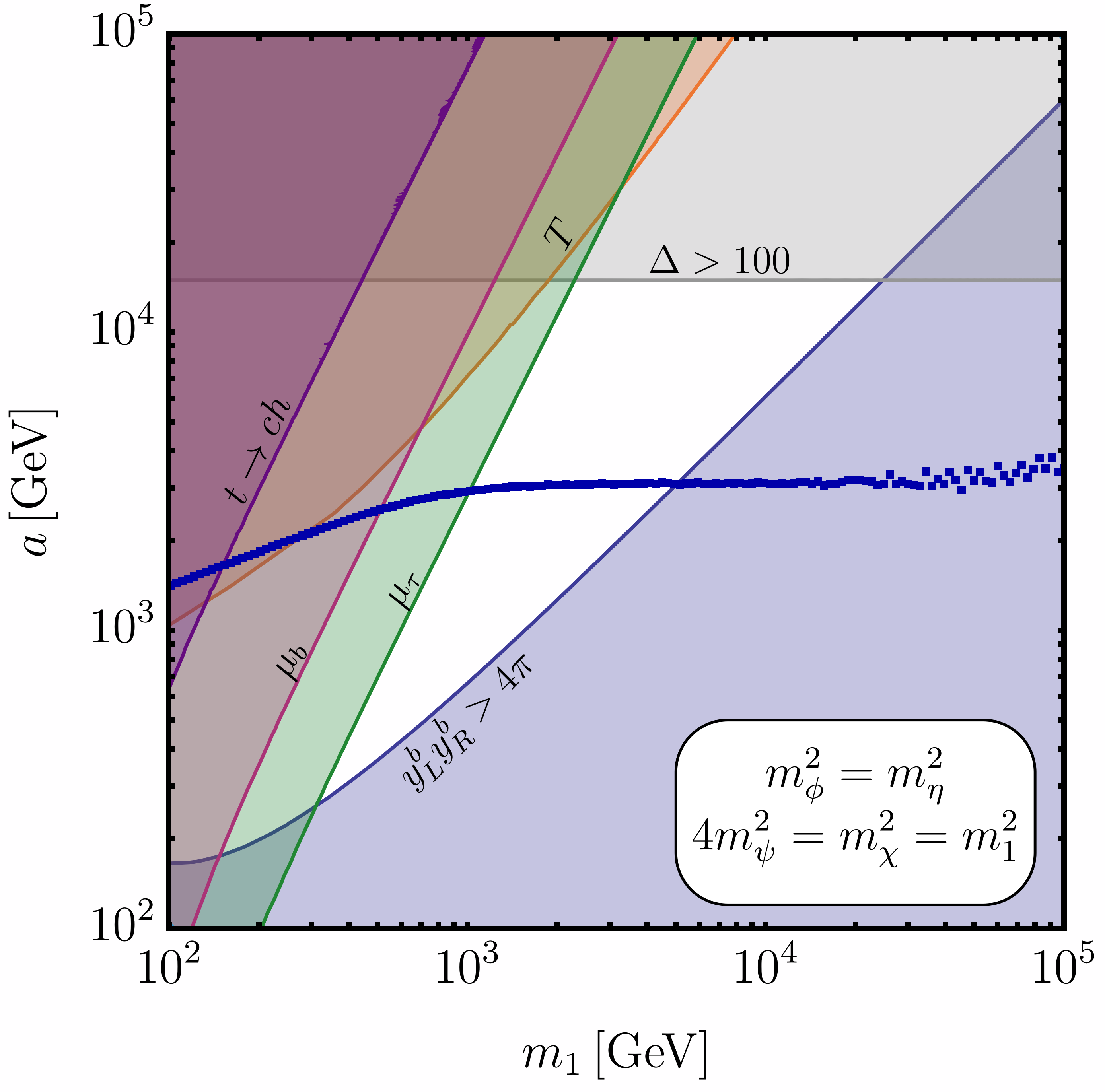}
    \caption{Exclusion plot with $\psi^{up}$ as a dark matter candidate. Blue dots are points for which the observed relic density is reproduced, given a set value for $m_\psi$.}
    \label{fig:darkmatter}
\end{figure}

Our model contains two dark matter candidates, the scalar $\varphi_1$ and the upper component of $\psi$. If $\psi^{up}$ is the lightest field, and thus a dark matter candidate, its relic abundance is set by (co)annihilation into $\tau$ and $\nu_L$. $\psi^{up}$ is pseudo-Dirac\footnote{It will receive a Majorana mass at one loop courtesy of $\nu_R$, however we expect this mass to be small given large $m_{\nu_R}$.}, so neglecting the masses of the SM fields allows us to write the thermally-averaged cross section as~\cite{diracdm}
\begin{equation}
    \expval{\sigma v} = \frac{m_\psi^2}{128\pi}  \left(\frac{4(y_L^L)^4 }{(m_\psi^2 + m_\phi^2)^2} +\frac{(y_R^\tau)^4}{(m_\psi^2 + m_\eta^2)^2}\right)~.
\end{equation}
We consider a freeze-out scenario, and perform our calculation before EWSB, assuming that this regime constitutes the majority of the temperature region between $m_\psi$ and the freeze-out temperature $T_f \sim m_\psi / 20$. Such an assumption is well-founded for $m_\psi$ above the TeV level, which turns out to be a highly relevant regime. Requiring that the correct dark matter relic density be reproduced gives the plot in Figure~\ref{fig:darkmatter}, with blue dots corresponding to parameter values for which this is accomplished. This shows that our model contains at least one viable dark matter candidate. A more thorough investigation would be required to fully examine all the possibilities.

\section{Future work}
\label{sec:futurework}

The most obvious and most important extension of this work is building a model treating all three generations of SM fermions, with lighter masses generated at progressively higher loop levels. This would require some additional symmetry structure, and possibly the introduction of more exotic fields to enable mass generation at higher loops. We note that for a straightforward inclusion of the fermions from other generations, assuming they are charged similarly under the new symmetries, the mass matrices for a fermion type will have the form
\begin{equation}
    M^{ij} = y_L^i y_R^j \times \mathrm{const.}
\end{equation}
This matrix is an outer product of two vectors and is thus rank one, implying only one fermion of each type can gain mass without further modifications to the theory.

A similar relic abundance analysis could be conducted for the other dark matter candidate, and the dark matter phenomenology for either candidate could also be explored further.

Finally, we note that it is fairly easy to include an axion in this model. If we introduce a scalar that transforms as $S \sim (\mathbf{1},\mathbf{1},0)(1)$ under $\mathcal{G}_{\mathrm{SM}} \otimes U(1)_\chi$, then the UV Lagrangian would include the term $S \overline{\chi_L} \chi_R$. When $S$ gets a VEV, the $U(1)_\chi$ symmetry would be spontaneously broken, resulting in an axion field, and the aforementioned term becoming the $\chi$ mass term.

\phantom{x}

\section{Conclusion}
\label{sec:conclusion}

We have extended an earlier framework to present a benchmark model that has one-loop-generated Dirac masses for the $\tau$, $b$, $c$, and $\nu_\tau$, thus explaining why $m_{\tau,b,c}$ are at the few-GeV level. We demonstrated that this model is currently viable in certain regions of parameter space, and showed how the parameter space will be further probed by future experiments, opening the prospect for discovery. 

Even if a discovery is not made, ensuing future constraints which maintain agreement with the SM will not be able to rule out such a model, only push the scale of new physics higher and necessitate a greater degree of fine-tuning, making the model less theoretically compelling. It is only in that sense that the SM tree-level fermion mass generation mechanism may become favoured. 

We briefly outlined avenues for future work in this area, including creating extended models that treat all three generations of SM fermions, exploring the phenomenology of dark matter candidates in the model, and/or introducing an axion to the theory. Note that a full three-family model will likely feature non-SM flavour effects for the lighter quarks and leptons, just as the presented model has the $t \to c h$ flavour-changing top quark decay. These processes are likely to be the best way to constrain complete radiative models. We expect, for example, that the masses of some exotics in some of these models will have to be so large to meet the flavour-violation constraints that consistency with electroweak naturalness will be impossible. A truly successful model would have to avoid this fate, and the search for such a model is a worthy goal of future research.

\begin{acknowledgments}
We thank Peter Cox for useful discussions. This work was supported in part by the Australian Research Council through the ARC Centre of Excellence for Dark Matter Particle Physics, CE200100008. LS is supported by an Australian Government Research Training Program Scholarship and the Rowden White Scholarship.
\end{acknowledgments}

\appendix

\bibliography{References}

@article{Baker_2021,
    author = "Baker, Michael J. and Cox, Peter and Volkas, Raymond R.",
    title = "{Has the Origin of the Third-Family Fermion Masses been Determined?}",
    eprint = "2012.10458",
    archivePrefix = "arXiv",
    primaryClass = "hep-ph",
    doi = "10.1007/JHEP04(2021)151",
    journal = "JHEP",
    volume = "04",
    pages = "151",
    year = "2021"
}

@article{Fraser_2014,
    author = "Fraser, Sean and Ma, Ernest",
    title = "{Anomalous Higgs Yukawa Couplings}",
    eprint = "1402.6415",
    archivePrefix = "arXiv",
    primaryClass = "hep-ph",
    reportNumber = "UCRHEP-T541-(FEB-2014), UCRHEP-T541-(MAR-2014), UCRHEP-T541-(JUL-2014), UCRHEP-T541-(SEP-2014)",
    doi = "10.1209/0295-5075/108/11002",
    journal = "EPL",
    volume = "108",
    number = "1",
    pages = "11002",
    year = "2014"
}

@phdthesis{Fraser:2016pmo,
    author = "Fraser, Sean Patrick",
    title = "{Anomalous Higgs Yukawa Couplings from Radiative Masses and Related Phenomena}",
    school = "UC, Riverside (main)",
    year = "2016"
}

@article{ParticleDataGroup:2024cfk,
    author = "Navas, S. and others",
    collaboration = "Particle Data Group",
    title = "{Review of particle physics}",
    doi = "10.1103/PhysRevD.110.030001",
    journal = "Phys. Rev. D",
    volume = "110",
    number = "3",
    pages = "030001",
    year = "2024"
}

@article{finetune,
    author = "Clarke, Jackson D. and Cox, Peter",
    title = "{Naturalness made easy: two-loop naturalness bounds on minimal SM extensions}",
    eprint = "1607.07446",
    archivePrefix = "arXiv",
    primaryClass = "hep-ph",
    doi = "10.1007/JHEP02(2017)129",
    journal = "JHEP",
    volume = "02",
    pages = "129",
    year = "2017"
}

@article{PTparams,
    author = "Peskin, Michael E. and Takeuchi, Tatsu",
    title = "{Estimation of oblique electroweak corrections}",
    reportNumber = "SLAC-PUB-5618",
    doi = "10.1103/PhysRevD.46.381",
    journal = "Phys. Rev. D",
    volume = "46",
    pages = "381--409",
    year = "1992"
}

@article{PTparamsearly,
    author = "Peskin, Michael E. and Takeuchi, Tatsu",
    title = "{New constraint on a strongly interacting Higgs sector}",
    reportNumber = "SLAC-PUB-5272",
    doi = "10.1103/PhysRevLett.65.964",
    journal = "Phys. Rev. Lett.",
    volume = "65",
    pages = "964--967",
    year = "1990"
}

@article{kappaframe,
    author = "David, A. and Denner, A. and Duehrssen, M. and Grazzini, M. and Grojean, C. and Passarino, G. and Schumacher, M. and Spira, M. and Weiglein, G. and Zanetti, M.",
    title = "{LHC HXSWG interim recommendations to explore the coupling structure of a Higgs-like particle}",
    journal = "",
    eprint = "1209.0040",
    archivePrefix = "arXiv",
    primaryClass = "hep-ph",
    reportNumber = "CERN-PH-TH-2012-284, LHCHXSWG-2012-001",
    month = "9",
    year = "2012"
}

@article{kappaframelong,
    author = "Andersen, J R and others",
    editor = "Heinemeyer, S and Mariotti, C and Passarino, G and Tanaka, R",
    collaboration = "LHC Higgs Cross Section Working Group",
    title = "{Handbook of LHC Higgs Cross Sections: 3. Higgs Properties}",
    journal = "",
    eprint = "1307.1347",
    archivePrefix = "arXiv",
    primaryClass = "hep-ph",
    reportNumber = "CERN-2013-004",
    doi = "10.5170/CERN-2013-004",
    month = "7",
    year = "2013"
}

@article{collider_future,
    author = "de Blas, J. and others",
    title = "{Higgs Boson Studies at Future Particle Colliders}",
    eprint = "1905.03764",
    archivePrefix = "arXiv",
    primaryClass = "hep-ph",
    reportNumber = "DESY-19-079",
    doi = "10.1007/JHEP01(2020)139",
    journal = "JHEP",
    volume = "01",
    pages = "139",
    year = "2020"
}

@article{STfuture,
    author = "de Blas, Jorge and Ciuchini, Marco and Franco, Enrico and Mishima, Satoshi and Pierini, Maurizio and Reina, Laura and Silvestrini, Luca",
    title = "{Electroweak precision observables and Higgs-boson signal strengths in the Standard Model and beyond: present and future}",
    eprint = "1608.01509",
    archivePrefix = "arXiv",
    primaryClass = "hep-ph",
    reportNumber = "KEK-TH-1919",
    doi = "10.1007/JHEP12(2016)135",
    journal = "JHEP",
    volume = "12",
    pages = "135",
    year = "2016"
}

@article{radiative1,
    author = "Weinberg, Steven",
    title = "{Electromagnetic and weak masses}",
    doi = "10.1103/PhysRevLett.29.388",
    journal = "Phys. Rev. Lett.",
    volume = "29",
    pages = "388--392",
    year = "1972"
}

@article{radiative2,
    author = "Balakrishna, B. S. and Kagan, A. L. and Mohapatra, R. N.",
    title = "{Quark Mixings and Mass Hierarchy From Radiative Corrections}",
    reportNumber = "MdDP-PP-88-142",
    doi = "10.1016/0370-2693(88)91676-0",
    journal = "Phys. Lett. B",
    volume = "205",
    pages = "345--352",
    year = "1988"
}

@article{radiative3,
    author = "Babu, K. S. and Ma, Ernest",
    title = "{Radiative Mechanisms for Generating Quark and Lepton Masses: Some Recent Developments}",
    reportNumber = "MdDP-PP-89-193, UCRHEP-T37",
    doi = "10.1142/S0217732389002239",
    journal = "Mod. Phys. Lett. A",
    volume = "4",
    pages = "1975",
    year = "1989"
}

@article{radiative5,
    author = "Ma, Ernest",
    title = "{Hierarchical Radiative Quark and Lepton Mass Matrices}",
    reportNumber = "UCRHEP-T53",
    doi = "10.1103/PhysRevLett.64.2866",
    journal = "Phys. Rev. Lett.",
    volume = "64",
    pages = "2866--2869",
    year = "1990"
}

@article{radiative4,
    author = "He, Xiao-Gang and Volkas, Raymond R. and Wu, Dan-Di",
    title = "{Radiative Generation of Quark and Lepton Mass Hierarchies From a Top Quark Mass Seed}",
    reportNumber = "UM-P-89/58, OZ-P-89/21",
    doi = "10.1103/PhysRevD.41.1630",
    journal = "Phys. Rev. D",
    volume = "41",
    pages = "1630",
    year = "1990"
}

@article{radiative6,
    author = "Dobrescu, Bogdan A. and Fox, Patrick J.",
    title = "{Quark and lepton masses from top loops}",
    eprint = "0805.0822",
    archivePrefix = "arXiv",
    primaryClass = "hep-ph",
    reportNumber = "FERMILAB-PUB-08-049-T",
    doi = "10.1088/1126-6708/2008/08/100",
    journal = "JHEP",
    volume = "08",
    pages = "100",
    year = "2008"
}

@article{radiative7,
    author = "Ma, Ernest",
    title = "{Radiative Origin of All Quark and Lepton Masses through Dark Matter with Flavor Symmetry}",
    eprint = "1311.3213",
    archivePrefix = "arXiv",
    primaryClass = "hep-ph",
    reportNumber = "UCRHEP-T538-(NOV-2013)",
    doi = "10.1103/PhysRevLett.112.091801",
    journal = "Phys. Rev. Lett.",
    volume = "112",
    pages = "091801",
    year = "2014"
}

@article{radiative8,
    author = "C\'arcamo Hern\'andez, A. E. and Kovalenko, Sergey and Schmidt, Ivan",
    title = "{Radiatively generated hierarchy of lepton and quark masses}",
    eprint = "1611.09797",
    archivePrefix = "arXiv",
    primaryClass = "hep-ph",
    doi = "10.1007/JHEP02(2017)125",
    journal = "JHEP",
    volume = "02",
    pages = "125",
    year = "2017"
}

@article{radiative9,
    author = "Arbel\'aez, Carolina and C\'arcamo Hern\'andez, A. E. and Cepedello, Ricardo and Kovalenko, Sergey and Schmidt, Ivan",
    title = "{Sequentially loop suppressed fermion masses from a single discrete symmetry}",
    eprint = "1911.02033",
    archivePrefix = "arXiv",
    primaryClass = "hep-ph",
    doi = "10.1007/JHEP06(2020)043",
    journal = "JHEP",
    volume = "06",
    pages = "043",
    year = "2020"
}

@article{radiative10,
    author = "Weinberg, Steven",
    title = "{Models of Lepton and Quark Masses}",
    eprint = "2001.06582",
    archivePrefix = "arXiv",
    primaryClass = "hep-th",
    reportNumber = "UTTG-10-19",
    doi = "10.1103/PhysRevD.101.035020",
    journal = "Phys. Rev. D",
    volume = "101",
    number = "3",
    pages = "035020",
    year = "2020"
}

@article{radiative11,
    author = "Mohanta, Gurucharan and Patel, Ketan M.",
    title = "{Radiatively generated fermion mass hierarchy from flavor nonuniversal gauge symmetries}",
    eprint = "2207.10407",
    archivePrefix = "arXiv",
    primaryClass = "hep-ph",
    doi = "10.1103/PhysRevD.106.075020",
    journal = "Phys. Rev. D",
    volume = "106",
    number = "7",
    pages = "075020",
    year = "2022"
}

@article{radiative12,
    author = "Bonilla, Cesar and Carcamo Hernandez, A. E. and Kovalenko, Sergey and Lee, H. and Pasechnik, R. and Schmidt, Ivan",
    title = "{Fermion mass hierarchy in an extended left-right symmetric model}",
    eprint = "2305.11967",
    archivePrefix = "arXiv",
    primaryClass = "hep-ph",
    doi = "10.1007/JHEP12(2023)075",
    journal = "JHEP",
    volume = "12",
    pages = "075",
    year = "2023"
}

@article{radiative13,
    author = "Jana, Sudip and Klett, Sophie and Lindner, Manfred and Mohapatra, Rabindra N.",
    title = "{Radiative origin of fermion mass hierarchy in left-right symmetric theory}",
    eprint = "2409.04246",
    archivePrefix = "arXiv",
    primaryClass = "hep-ph",
    doi = "10.1007/JHEP01(2025)082",
    journal = "JHEP",
    volume = "01",
    pages = "082",
    year = "2025"
}

@article{radiative14,
    author = "Baker, Michael J. and Cox, Peter and Volkas, Raymond R.",
    title = "{Radiative muon mass models and $(g-2)_\mu$}",
    eprint = "2103.13401",
    archivePrefix = "arXiv",
    primaryClass = "hep-ph",
    doi = "10.1007/JHEP05(2021)174",
    journal = "JHEP",
    volume = "05",
    pages = "174",
    year = "2021"
}

@article{radiative15,
    author = "Crivellin, Andreas and Hofer, Lars and Nierste, Ulrich and Scherer, Dominik",
    title = "{Phenomenological consequences of radiative flavor violation in the MSSM}",
    eprint = "1105.2818",
    archivePrefix = "arXiv",
    primaryClass = "hep-ph",
    doi = "10.1103/PhysRevD.84.035030",
    journal = "Phys. Rev. D",
    volume = "84",
    pages = "035030",
    year = "2011"
}

@article{ESUbriefing,
    author = "Ellis, Richard Keith and others",
    title = "Physics Briefing Book: Input for the European Strategy for Particle Physics Update 2020",
    journal = "",
    eprint = "1910.11775",
    archivePrefix = "arXiv",
    primaryClass = "hep-ex",
    reportNumber = "CERN-ESU-004",
    month = "10",
    year = "2019"
}

@article{tchbound,
    author = "Aad, Georges and others",
    collaboration = "ATLAS",
    title = "{Search for flavour-changing neutral-current couplings between the top quark and the Higgs boson in multi-lepton final states in 13~TeV pp collisions with the ATLAS detector}",
    eprint = "2404.02123",
    archivePrefix = "arXiv",
    primaryClass = "hep-ex",
    reportNumber = "CERN-EP-2024-070, CERN-EP-2024-070",
    doi = "10.1140/epjc/s10052-024-12994-1",
    journal = "Eur. Phys. J. C",
    volume = "84",
    number = "7",
    pages = "757",
    year = "2024"
}

@article{CMS_cc,
    author = "Hayrapetyan, Aram and others",
    collaboration = "CMS",
    title = "{Simultaneous Probe of the Charm and Bottom Quark Yukawa Couplings Using tt{\textasciimacron}H Events}",
    eprint = "2509.22535",
    archivePrefix = "arXiv",
    primaryClass = "hep-ex",
    reportNumber = "CMS-HIG-24-018, CERN-EP-2025-202",
    doi = "10.1103/9nwb-splk",
    journal = "Phys. Rev. Lett.",
    volume = "136",
    number = "1",
    pages = "011801",
    year = "2026"
}

@article{CMS_sum,
    author = "Tumasyan, Armen and others",
    collaboration = "CMS",
    title = "{A portrait of the Higgs boson by the CMS experiment ten years after the discovery.}",
    eprint = "2207.00043",
    archivePrefix = "arXiv",
    primaryClass = "hep-ex",
    reportNumber = "CMS-HIG-22-001, CERN-EP-2022-039",
    doi = "10.1038/s41586-022-04892-x",
    journal = "Nature",
    volume = "607",
    number = "7917",
    pages = "60--68",
    year = "2022",
    note = "[Erratum: Nature 623, (2023)]"
}

@article{ATLAS_bb,
    author = "Aad, Georges and others",
    collaboration = "ATLAS",
    title = "{Measurements of WH and ZH production with Higgs boson decays into bottom quarks and direct constraints on the charm Yukawa coupling in 13 TeV pp collisions with the ATLAS detector}",
    eprint = "2410.19611",
    archivePrefix = "arXiv",
    primaryClass = "hep-ex",
    reportNumber = "CERN-EP-2024-237",
    doi = "10.1007/JHEP04(2025)075",
    journal = "JHEP",
    volume = "04",
    pages = "075",
    year = "2025"
}

@article{ATLAS_tau,
    author = "Aaboud, Morad and others",
    collaboration = "ATLAS",
    title = "{Cross-section measurements of the Higgs boson decaying into a pair of $\tau$-leptons in proton-proton collisions at $\sqrt{s}=13$ TeV with the ATLAS detector}",
    eprint = "1811.08856",
    archivePrefix = "arXiv",
    primaryClass = "hep-ex",
    reportNumber = "CERN-EP-2018-232",
    doi = "10.1103/PhysRevD.99.072001",
    journal = "Phys. Rev. D",
    volume = "99",
    pages = "072001",
    year = "2019"
}

@article{diracdm,
    author = "Harnik, Roni and Kribs, Graham D.",
    title = "{An Effective Theory of Dirac Dark Matter}",
    eprint = "0810.5557",
    archivePrefix = "arXiv",
    primaryClass = "hep-ph",
    reportNumber = "SLAC-PUB-13482",
    doi = "10.1103/PhysRevD.79.095007",
    journal = "Phys. Rev. D",
    volume = "79",
    pages = "095007",
    year = "2009"
}

@article{VLQs,
    author = "Banerjee, Avik and Bergeaas Kuutmann, Elin and Ellajosyula, Venugopal and Enberg, Rikard and Ferretti, Gabriele and Panizzi, Luca",
    title = "{Vector-like quarks: Status and new directions at the LHC}",
    eprint = "2406.09193",
    archivePrefix = "arXiv",
    primaryClass = "hep-ph",
    doi = "10.21468/SciPostPhysCore.7.4.079",
    journal = "SciPost Phys. Core",
    volume = "7",
    pages = "079",
    year = "2024"
}

@article{VLQrev,
    author = "Benbrik, Rachid and Boukidi, Mohammed and Ech-chaouy, Mohamed and Moretti, Stefano and Salime, Khawla and Yan, Qi-Shu",
    title = "{Vector-Like Quarks at the LHC: A unified perspective from ATLAS and CMS exclusion limits}",
    eprint = "2412.01761",
    archivePrefix = "arXiv",
    primaryClass = "hep-ph",
    doi = "10.1007/JHEP03(2025)020",
    journal = "JHEP",
    volume = "03",
    pages = "020",
    year = "2025"
}

@article{squark1,
    author = "Sirunyan, Albert M and others",
    collaboration = "CMS",
    title = "{Searches for physics beyond the standard model with the $M_\mathrm{T2}$ variable in hadronic final states with and without disappearing tracks in proton-proton collisions at $\sqrt{s}=$ 13 TeV}",
    eprint = "1909.03460",
    archivePrefix = "arXiv",
    primaryClass = "hep-ex",
    reportNumber = "CMS-SUS-19-005, CERN-EP-2019-180",
    doi = "10.1140/epjc/s10052-019-7493-x",
    journal = "Eur. Phys. J. C",
    volume = "80",
    number = "1",
    pages = "3",
    year = "2020"
}

@article{squark2,
    author = "Aad, Georges and others",
    collaboration = "ATLAS",
    title = "{Search for new phenomena in events with an energetic jet and missing transverse momentum in $pp$ collisions at $\sqrt {s}$ =13  TeV with the ATLAS detector}",
    eprint = "2102.10874",
    archivePrefix = "arXiv",
    primaryClass = "hep-ex",
    reportNumber = "CERN-EP-2020-238",
    doi = "10.1103/PhysRevD.103.112006",
    journal = "Phys. Rev. D",
    volume = "103",
    number = "11",
    pages = "112006",
    year = "2021"
}

@article{SUSYbound,
    author = "Aad, Georges and others",
    collaboration = "ATLAS",
    title = "{Search for electroweak production of supersymmetric particles in final states with two {\ensuremath{\tau}}-leptons in $ \sqrt{s} $ = 13 TeV pp collisions with the ATLAS detector}",
    eprint = "2402.00603",
    archivePrefix = "arXiv",
    primaryClass = "hep-ex",
    reportNumber = "CERN-EP-2023-295",
    doi = "10.1007/JHEP05(2024)150",
    journal = "JHEP",
    volume = "05",
    pages = "150",
    year = "2024"
}

@PREAMBLE{
 "\providecommand{\noopsort}[1]{}" 
 # "\providecommand{\singleletter}[1]{#1}%" 
}

\end{document}